\documentstyle[emulateapj,epsfig]{article}

\slugcomment{}

\begin{document}

\title{
Forming the First Stars in the Universe: The
Fragmentation of\\ Primordial Gas}

\author{Volker Bromm, Paolo S. Coppi, and Richard B. Larson}
\affil{Department of Astronomy, Yale University, New Haven, CT 06520-8101;\\
volker@astro.yale.edu, coppi@astro.yale.edu, larson@astro.yale.edu}

\begin{abstract}

In order to constrain the initial mass function (IMF) of the first generation
of stars (Population III), we investigate the fragmentation properties
of metal-free gas in the context of a hierarchical model of structure
formation. We investigate the evolution of an isolated 3 $\sigma$ peak
of mass $2\times 10^{6}M_{\odot}$ which collapses at $z_{coll}\simeq 30$
using Smoothed Particle Hydrodynamics.
We find that the gas dissipatively settles into a rotationally supported disk
which has a very filamentary morphology.
The gas in these filaments is Jeans unstable with
$M_{J}\sim 10^{3}M_{\odot}$. Fragmentation leads to the formation
of high density ($n>10^{8}$ cm$^{-3}$) clumps which subsequently grow
in mass by accreting surrounding gas and by merging with other clumps up to
masses of $\sim 10^{4}M_{\odot}$. This suggests that the very first stars
were rather massive.
We explore the complex dynamics of the 
merging and tidal disruption of these clumps by following
their evolution over a few dynamical times.
\end{abstract}
\keywords{cosmology: theory --- early universe --- galaxies: formation
--- hydrodynamics}

\section{INTRODUCTION}

Little is known about the history of the universe at redshifts 
$z\simeq 1000-5$, corresponding to $10^{6}-10^{9}$ years after the Big Bang,
and only recently have astrophysicists begun to seriously investigate this 
crucial post-recombination era. At $z\sim 1000$, the photons of the
cosmic microwave background (CMB) shifted into the infrared, and the
universe entered what has been termed the ``Dark Ages'' (Rees 1999).
We know that the universe was reionized again before a redshift of $\sim 5$ from
the absence of Gunn-Peterson absorption in the spectra of high redshift quasars.
One of the grand challenges in modern cosmology is to determine
when the universe was lit up again by the first luminous objects (the so-called
Population III) which were responsible for reionizing and reheating the
intergalactic medium 
(Haiman \& Loeb 1997; Ferrara 1998).
The very first generation of stars must
have formed out of probably unmagnetized, pure H/He gas, since
heavy elements can only be produced
in the interior of stars. 
These characteristics render the primordial star formation problem very
different from the present-day case, and lead to a significant simplification
of the relevant physics (Larson 1998; Loeb 1998).

To determine the characteristic mass scale of the Population III stars, and to
constrain their initial mass function (IMF), one has to study the collapse and
fragmentation of metal-free gas. This problem has been addressed by a number of
authors (e.g., Yoneyama 1972; Hutchins 1976; Carlberg 1981; Kashlinsky \& Rees 1983;
Palla, Salpeter, \& Stahler 1983;
Silk 1983;
Haiman, Thoul, \& Loeb 1996; Uehara et al. 1996;
Tegmark et al. 1997;
Omukai \& Nishi 1998;
Nakamura \& Umemura 1999).
Recently, three-dimensional 
cosmological simulations have reached sufficient resolution on very
small scales to approach the Population III star formation problem 
(Anninos \& Norman 1996; Ostriker \& Gnedin 1996; Abel et al. 1998a).

Complementary to these last studies, we explore
the fragmentation of primordial gas under a variety
of initial conditions.
Our method allows us to follow the evolution of the
fragments for
a few dynamical times, and to include the physics of competitive accretion.
These capabilities are crucial for the investigation of the
resulting mass spectrum. 
A more detailed exposition of our code, together with further results 
of our exploratory survey, will be presented in
a forthcoming publication (Bromm, Coppi, \& Larson 1999). 
In this Letter, we focus on one representative experiment
which addresses the key
question: {\it How does the fragmentation of the first collapsing gas clouds
proceed in a hierarchical (bottom up) scenario of structure formation?}

\section{NUMERICAL METHOD}

Our code is based on a version
of TREESPH (Hernquist \& Katz 1989) which combines
the Smoothed Particle Hydrodynamics (SPH) method (e.g., Monaghan 1992) with a hierarchical (tree) gravity solver.
To study primordial gas, we have made a number of additions.
Most importantly, radiative cooling due to hydrogen molecules
has been taken into account. In the absence of metals, H$_{2}$ is the
main coolant below $\sim 10^{4}$ K, the typical
temperature range in collapsing Population III objects. We implement 
the H$_{2}$ cooling
function given by Galli \& Palla (1998).
At temperatures approaching
$10^{4}$ K, cooling due to lines of atomic hydrogen dominates (Katz
\& Gunn 1991).
The efficiency of H$_{2}$ cooling is very sensitive to the H$_{2}$ abundance.
Therefore, it is necessary to compute the nonequilibrium evolution
of the primordial chemistry. We take into account the six species
H, H$^{+}$, H$^{-}$, H$_{2}$, H$_{2}^{+}$, and e$^{-}$. Helium is
neglected, since the temperatures in our application are low enough
($<10^{4}$ K) to render the He species almost inert. Reaction
rates are taken from Haiman et al. (1996).
In addition, we have included 3-body reactions which become important at
high density ($n > 10^{8}$ cm$^{-3}$), and can convert the gas
almost completely into molecular form (Palla et al. 1983).
The chemistry network
is solved by 
the approximate backwards differencing formula (BDF) method 
(Anninos et al. 1997).

We have devised an algorithm to merge SPH particles
in high density regions to overcome the otherwise prohibitive
timestep limitation, as enforced
by the Courant stability criterion (Bate, Bonnell, \& Price 1995).
To follow the simulation for a few dynamical times, we 
allow SPH particles to merge into more
massive ones, provided they
exceed a pre-determined density threshold, typically $10^{8}-10^{10}$ cm$^{-3}$.
More details of the merging algorithm will be given in Bromm et al. (1999).

\section{THE SIMULATION}
\subsection{Initial Conditions}
 
Within a hierarchical cosmogony, the
very first stars are expected to form out
of $3-4 \sigma$ peaks in the random field of primordial density fluctuations.
The early (linear) evolution of such a peak, assumed to be an isolated and
roughly spherical overdensity, can be described by the top-hat
model (e.g., Padmanabhan 1993). We use the top-hat approximation to
specify the initial dark matter (DM) configuration, where we choose
the background universe to be described by the critical Einstein-de Sitter
solution with density parameters: $\Omega_{DM}=0.95$, $\Omega_{B}=0.05$, and 
Hubble constant $h=H_{0}/(100$ km s$^{-1}$ Mpc$^{-1}$)=0.5.
In this paper, we investigate the fate of a $3 \sigma$ peak
of total mass $2\times 10^{6}M_{\odot}$, corresponding to $10^{5}M_{\odot}$
in baryons which is
close to the
cosmological Jeans mass (Couchman \& Rees 1986).
On this mass scale, the
standard Cold Dark Matter (CDM) scenario predicts a present-day r.m.s. 
overdensity of $\sigma_{0}(M)\simeq 16$, with a normalization
$\sigma_{8}=1$ on the $8h^{-1}$Mpc scale. Then, one can estimate the
redshift of collapse (or virialization) from:
$1+z_{coll}=3 \sigma_{0}(M)/1.69$, leading to $z_{coll}\simeq 30$.
Our simulation is initialized at $z_{i}=100$, by performing the following steps.

The collisionless DM particles are placed on
a cubical Cartesian grid, and are then perturbed according
to a given power spectrum $P(k)=A k^{n}$, by applying the 
Zel'dovich approximation (Zel'dovich 1970) which 
also allows to self-consistently
assign initial velocities. The power-law index is set to $n=-3$ which
is the asymptotic small-scale 
behavior
in the standard CDM model (Peebles 1993). The amplitude $A$ is adjusted, so that
the fundamental mode (i.e., the smallest contributing wavenumber $k_{min}$),
has an initial mean square overdensity which grows to $P(k_{min})\simeq 1$
at collapse. Next, particles within a (proper) radius of $R_{i}=$ 150 pc 
are selected for the
simulation. The resulting number of DM particles is $N_{DM}=14123$.
Finally, the particles are set into rigid rotation and are endowed
with a uniform Hubble expansion
(see also Katz 1991).
Angular momentum is added by assuming a spin-parameter
$\lambda=L|E|^{1/2}/(G M^{5/2})=0.05$,
where $L$, $E$, and $M$ are the total angular momentum, energy, and mass,
respectively.

The collisional SPH particles ($N_{SPH}=16384$ in this simulation) are randomly
placed to approximate
a uniform initial density. The random sampling inevitably introduces
shot-noise. 
The DM particles were set up on a regular grid specifically to avoid the
unphysical shot-noise distribution which could mask the desired
physical power spectrum. For the gas, however, the presence
of the shot noise is not a big problem, since the gas mass is initially
smaller than the Jeans mass. Therefore, sound waves will efficiently wipe out
all initial density disturbances.
The SPH particles
are endowed with the same Hubble expansion and rigid rotation as the DM ones.
Since $z<100$, heating, cooling, and photoreactions due to the CMB can be
safely neglected here.
For the chemical abundances and the gas temperature,
we adopt the 
initial values (see Tegmark et al. 1997): 
$x_{i}=4.6\times10^{-4}$, $f_{i}=10^{-4}$, and
$T_{gas, i}\simeq 200$ K, where $x_{i}$, $f_{i}$ are the initial free electron
and H$_{2}$ abundances, respectively.
The use of a more realistic rate for the photodissociation of H$_{2}^{+}$
predicts a lower initial H$_{2}$ abundance, typically $f_{i}\simeq 10^{-6}$.
The evolution of a simulation initialized with such a lower value, however,
quickly converges to the case presented in this paper.

\subsection{Results}
\subsubsection{\it Free-fall Phase}

From turnaround ($z=50$) to the moment of 
collapse ($z\simeq 30$), both the DM and the gas are freely falling.
In response to the initially imprinted 
$P(k)\propto k^{-3}$ noise, the DM develops marked substructure.
This happens in a hierarchical fashion, where smaller clumps
form first, and rapidly merge into larger aggregations.
The gas initially falls together with the DM, heating up
adiabatically, $T\propto n^{2/3}$ for $\gamma = 5/3$, until the temperature 
reaches the virial value of $T_{vir}\sim$ 5000 K which is determined by the
depth of the DM potential well.
The virialized gas is able to cool efficiently, and 
the temperature decreases again with continuing compression
down to a few 100 K.
At first, the gas is not able to condense onto the shallow DM potential wells,
and its mass distribution remains smooth. In the course of the collapse,
however, the baryonic Jeans mass decreases enough to eventually allow
the gas to fall into the deepening DM troughs. 
This
gravitational ``head-start'' is important because it determines 
where fragmentation occurs first, resulting in the
most massive gas clumps.

Briefly after collapse, all the substructure in the DM has been
wiped out in the process of violent relaxation (Lynden-Bell 1967) which
is very effective in quickly establishing virial equilibrium on the
dynamical timescale of the DM halo ($\sim 10^{7}$ years).
Before having been wiped out, however,
the DM substructure has left its imprint on the gas, determining the 
morphology of the incipient disk. 
The DM
reaches an equilibrium configuration with
a core radius of $\sim$ 10 pc and an approximately isothermal
density profile ($\rho \propto r^{-2}$) further out.

The central gas disk is horizontally supported by rotation. In the
presence of a DM halo, one expects a contraction by a factor of
$1/\lambda\simeq 20$ for rotational support. The dimension of the
disk, as shown in Figure 2,
is in good accordance with this prediction. Vertically, the disk is
roughly pressure supported with a scale height of $H=c_{s}^{2}/(G\Sigma)\simeq$
2 pc, where $c_{s}\simeq$ 2 km/s is the sound speed, and $\Sigma$ the
gas surface density. The resulting
disk has a very filamentary and knotty structure with typical
densities of $n\sim 10^{4}$ cm$^{-3}$ and temperatures of $T\sim 500$ K,
corresponding to a Jeans mass of $M_{J}\sim 10^{3}M_{\odot}$. 
We next discuss the fragmentation of these filamentary features.

\subsubsection{\it Disk-like Phase}

The gas which makes up the filaments is Jeans unstable
since $t_{ff}<t_{sound}$, where the free-fall time is
$t_{ff}\simeq (G m_{\mbox{\scriptsize H}} n)^{-1/2}$, and the sound-crossing
time $t_{sound}\simeq R/c_{s}$. $R\sim$ 15 pc is the
radius of the disk. In Figure 1, 
the free-fall time is compared to the sound-crossing time and to the cooling time, 
$t_{cool}\simeq n k_{B} T/(f n^{2} L)$.
Here, $f$ is the fractional H$_{2}$ abundance,
and $L$ the cooling rate (in erg s$^{-1}$ cm$^{3}$).
For densities $n> 10^{3}$ cm$^{-3}$, the gas becomes Jeans unstable.
The onset of instability
roughly coincides with the condition that $t_{cool}\simeq t_{ff}$.
The value $n\sim 10^{3}$ cm$^{-3}$ is also the critical density for
the rotational-vibrational levels of H$_{2}$ being populated according
to local thermodynamic equilibrium (LTE).

The first regions to undergo runaway collapse lead to the
formation of dense ($n>10^{8}$ cm$^{-3}$) gas clumps of mass $\sim 10^{3}M_{\odot}$.
The dynamical state of a typical clump can be described by the ratios
$E_{th}/|E_{grav}|\simeq 0.5$, and $E_{rot}/|E_{grav}|\simeq 0.1$, where
$E_{th}$, $E_{rot}$, $E_{grav}$ are the thermal, rotational, and
gravitational energies of the clump.
The further
evolution of the disk consists of a complex history of new clumps forming,
and old ones continually accreting gas and occasionally merging with
other clumps. 
The outcome of this
evolution is shown in Figure 2, corresponding to the
situation at $z=28$. The disk-like structure has fragmented into
13 clumps, ranging in mass from 220 to $\sim 10^{4}M_{\odot}$.
At this stage,
$\sim$50\% of the total gas mass has been incorporated into high density
clumps. 
This fraction is unrealistically high due to the homologous
nature of the top-hat collapse where all particles reach the center at the same
time. Abel et al. (1998a), on the other hand, estimate this fraction
to be $\sim$8\%. This value is more realistic, since these authors
consider an overdensity which is initially already centrally concentrated,
leading to a collapse which is spread out in time.
Since the larger clumps tend to retain their individuality, the final
evolutionary state of the disk is expected to be a small cluster of
isolated clumps.

Figure 3 summarizes the thermodynamic state of the gas together with the
chemical abundances. 
When a region undergoes runaway collapse,
particles are drawn from the quasi-stationary ``reservoir'' at
$n\sim 10^{4}$ cm$^{-3}$ which corresponds to the filamentary gas, and
swiftly move to higher densities at
roughly constant temperature (Fig. 3c).
At a density of $n=10^{8}$cm$^{-3}$,
SPH particles are merged into more massive ones, as described in \S 2.
At higher densities, the Jeans mass would decrease below the resolution
limit ($M_{res}\sim 200 M_{\odot}$; Bate \& Burkert 1997) of our
simulation (Fig. 3d).
In order to reliably follow the evolution of
the gas to even higher densities, a larger number of SPH particles
would be needed. In this case, one would expect the gas density to
increase up to the point where the now fully molecular gas
becomes opaque, and subsequently
behaves adiabatically again.
The present study, however, has the purpose of following the evolution
of many gas clumps for a few dynamical times. Therefore, merging at
some point is inevitable.

\section{SUMMARY AND CONCLUSIONS}

We have discussed the evolution of a primordial star forming region
in the context of the standard CDM model for structure formation.
In the early stages of the collapse, the morphology of the DM substructure
imprints a pattern onto the initially completely smooth gas which
determines where gas fragmentation is to occur first.
Therefore, the initial power spectrum of the DM fluctuations might be 
an important ingredient for primordial star formation.
Simultaneously with the virialization of the DM component, the
gas dissipatively settles down into a compact central disk. This disk has a
very filamentary structure, and
is Jeans unstable which leads to fragmentation 
into clumps of initially $\sim 10^{2}-
10^{3}M_{\odot}$. These clumps grow in mass by both accreting surrounding gas,
and by merging with other high-density clumps. At the end of this process,
a small number ($\sim$ 10) of these clumps have formed.

Population III stars, then, might have been quite massive, perhaps
even `very massive' ($M_{char}>100 M_{\odot}$), if one assumes that 
a Jeans mass clump does not undergo significant further fragmentation.
Although subsequent
fragmentation cannot be excluded, this is expected to
lead to the formation of a binary or a small multiple system, if experience
from present-day star formation offers any guidance here (Larson 1998).
Our results agree well
with those of Abel, Bryan, \& Norman (1998b) who have applied
the Adaptive Mesh Refinement (AMR) method to the problem. These authors
find a dense core of order a few times $10^{2}M_{\odot}$, which continues
to accrete mass and shows no sign of further subfragmentation.
Clearly, high resolution studies of the further clump evolution are
necessary, including a treatment of opacity effects (see Omukai \&
Nishi 1998). We are presently engaged in doing so and will present
our findings in a future publication.

The effects of
cooling due to HD molecules (Galli \& Palla 1998),
and of the presence of the CMB on the H$_{2}$ level populations 
(Capuzzo-Dolcetta, Di Fazio, \& Palla 1991), were not
included in our work, and both could affect the estimate of the Jeans mass.
Test calculations indicate that the effect of HD cooling may not significantly
change our results, but a more complete treatment will be incorporated
in subsequent work.


\acknowledgments{We would like to thank Z. Haiman and A. Loeb for
providing us with their chemical reaction rates and for helpful
discussions, L. Hernquist for making available to us a version of TREESPH, and
T. Abel, A. Ferrara, and F. Palla for detailed comments.
Support from the NASA ATP grant NAG5-7074 is gratefully acknowledged.
}


\clearpage

\begin{center}
\epsfig{file=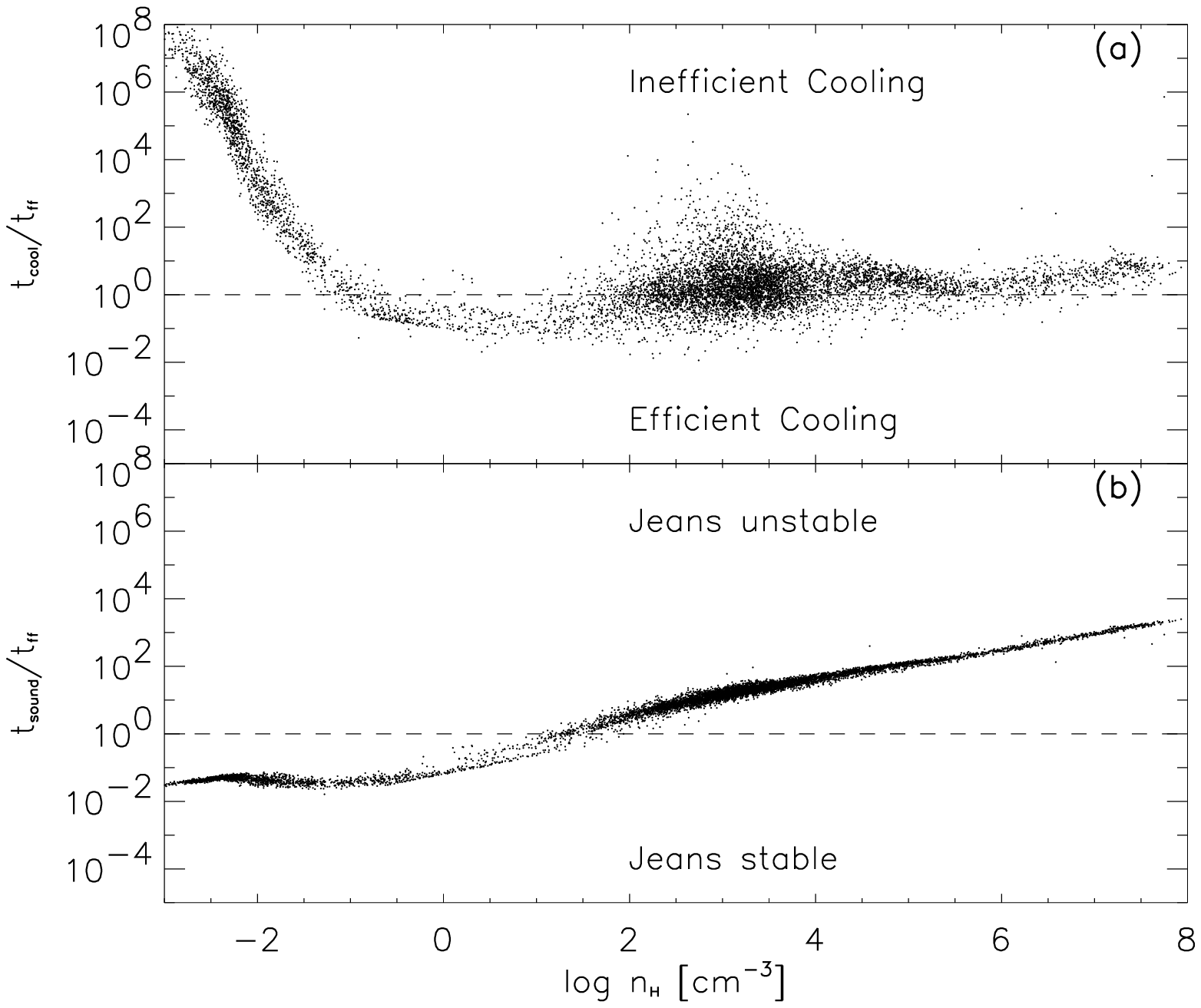,width=14cm,height=12.6cm}
\figcaption{
Important timescales at $z=28$.
{\bf (a)} Ratio of cooling and free-fall timescales vs. gas density.
At low density, cooling is not efficient and contraction proceeds 
adiabatically. For higher densities, characteristic of the gas
after virialization, cooling is very efficient up to densities
$n\sim 10^{3}$ cm$^{-3}$.
The evolution from $n\sim 10^{3}$ cm$^{-3}$ to 
$n\sim 10^{8}$ cm$^{-3}$ proceeds such that $t_{cool}\simeq t_{ff}$.
{\bf (b)} Ratio of sound-crossing and free-fall timescales vs.
gas density. For $t_{ff}<t_{sound}$, the gas is Jeans unstable. 
Gravitational instabilty sets in at $n\sim 10^{3}$ cm$^{-3}$ which corresponds
to the compressed gas in the disk-like configuration.
The onset of instability roughly coincides with
the condition that $t_{cool}\simeq t_{ff}$ in panel (a).
\label{fig1}}
\end{center}

\begin{center}
\epsfig{file=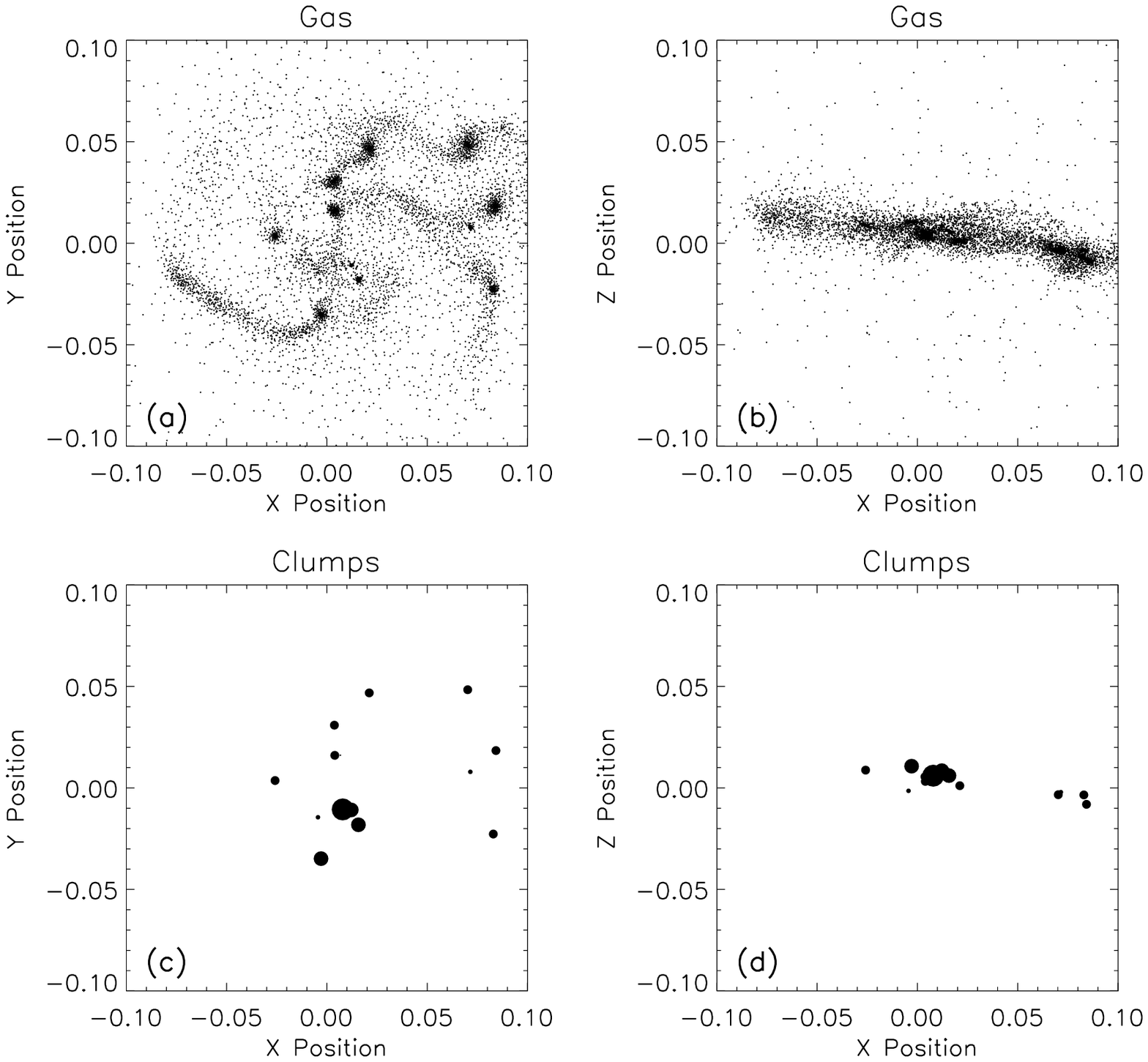,width=14cm,height=12.6cm}

\figcaption{Situation at $z=28$.
Shown is the ongoing fragmentation of the gas disk in the center of the
DM halo.
The (proper) length of the box is 30 pc.
{\it Left Panels:} Face-on view. {\it Right Panels:} Edge-on view.
{\bf (a)-(b)} Gas component. The gas which had been assembled in a
disk-like configuration, has undergone further fragmentation into roughly
spherical high-density ($n>10^{8}$ cm$^{-3}$) clumps.
{\bf (c)-(d)} Clump distribution. The four increasing dot sizes denote
increasing mass 
scale: $10^{2}-10^{3}M_{\odot}$,
$10^{3}-5\times 10^{3}M_{\odot}$,
$5\times 10^{3}-10^{4}M_{\odot}$, and
$M>10^{4}M_{\odot}$. By now, $\sim$ 50 \% of the total gas
mass is incorporated into high-density clumps. The most massive clumps
have already accreted all of the surrounding
gas. This explains the paucity of unmerged gas particles
close to the center of the disk in panel (a).
\label{fig2}}
\end{center}

\begin{center}
\epsfig{file=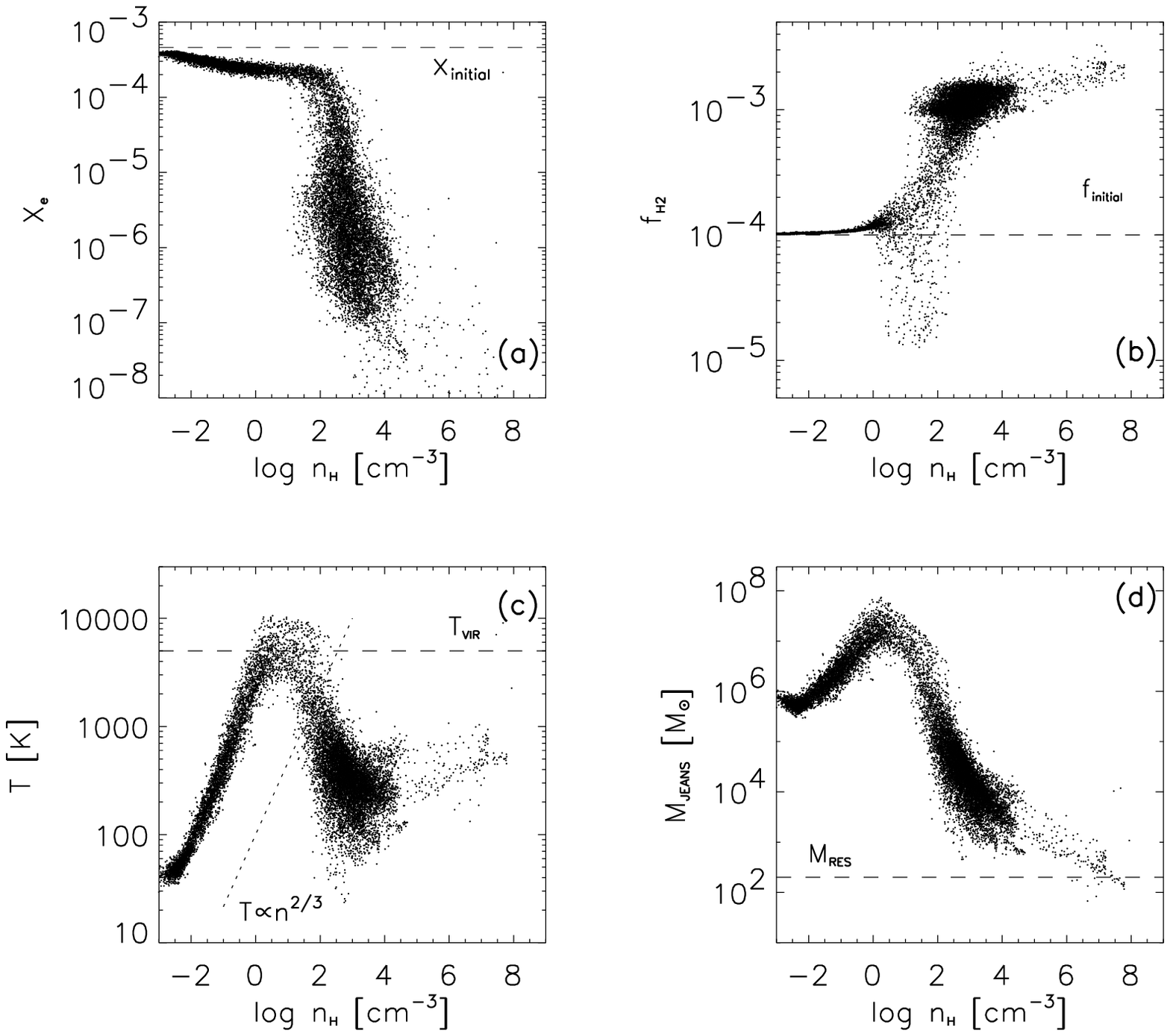,width=14cm,height=12.6cm}
\figcaption{Gas properties at $z=30.5$.
{\bf (a)} Free electron abundance vs. gas number density. At high
density, the residual electrons recombine until the gas is almost completely
neutral (at $n\sim 10^{4}$ cm$^{-3}$). {\bf (b)} H$_{2}$ abundance
vs. gas number density. Due to the H$^{-}$ channel, the asymptotic value
of $\sim 10^{-3}$ has been reached for densities $n> 10^{3}$ cm$^{-3}$.
{\bf (c)} Gas temperature vs. number density. At low gas densities,
the temperature rises due to adiabatic compression until it reaches
the virial value. At higher densities, H$_{2}$ line cooling drives the
temperature down again until the gas settles into a quasi-equilibrium state
at $T\simeq$ 500 K and $n\simeq 10^{4}$ cm$^{-3}$.
{\bf (d)} Jeans mass vs. number density. 
The Jeans mass 
reaches a value of $M_{J}\simeq 10^{3}M_{\odot}$ in
the disk-like central feature, and approaches
the resolution limit of the simulation, $M_{res}\sim 200 M_{\odot}$, for
densities close to the merging threshold of $n=10{^8}$ cm$^{-3}$.
\label{fig3}}
\end{center}
\end{document}